\documentclass[conference]{IEEEtran}
\IEEEoverridecommandlockouts
\usepackage{cite}
\usepackage{amsmath,amssymb,amsfonts}
\usepackage{algorithmic}
\usepackage{graphicx}
\usepackage{textcomp}
\usepackage{xcolor}
\usepackage{multirow}
\usepackage[symbol]{footmisc}
\usepackage{booktabs}
\usepackage{tabularx}
\usepackage{multirow}
\usepackage{graphicx}
\usepackage{xcolor}
\usepackage{colortbl}
\usepackage[table]{xcolor}
\usepackage{float}
\usepackage[colorinlistoftodos]{todonotes}

\definecolor{headblue}{RGB}{218, 232, 252}  
\definecolor{cellgreen}{RGB}{213, 232, 212} 
\definecolor{cellgray}{RGB}{220, 220, 220}  

\def\BibTeX{{\rm B\kern-.05em{\sc i\kern-.025em b}\kern-.08em
    T\kern-.1667em\lower.7ex\hbox{E}\kern-.125emX}}
\begin{document}

\newcommand\copyrighttext{%
  \footnotesize \textcopyright 2026 IEEE. Personal use of this material is permitted. Permission from IEEE must be obtained for all other uses, in any current or future media, including reprinting/republishing this material for advertising or promotional purposes, creating new collective works, for resale or redistribution to servers or lists, or reuse of any copyrighted component of this work in other works. 
  \linebreak Preprint submitted to the 12th IEEE International Smart Cities Conference 2026 (ISC2 2026).
  }
\newcommand\copyrightnotice{%
\begin{tikzpicture}[remember picture,overlay]
\node[anchor=south,yshift=10pt] at (current page.south) {\fbox{\parbox{\dimexpr\textwidth-\fboxsep-\fboxrule\relax}{\copyrighttext}}};
\end{tikzpicture}%
}

\title{From Sensor Data to Classroom Inquiry: GenAI-Supported Exploration of School Digital Twin Data}
\makeatletter
\newcommand{\linebreakand}{%
  \end{@IEEEauthorhalign}
  \hfill\mbox{}\par
  \mbox{}\hfill\begin{@IEEEauthorhalign}
}
\makeatother

\author{
\IEEEauthorblockN{Themistoklis Sarantakos}
\IEEEauthorblockA{\textit{Industrial Systems Institute} \\
\textit{Athena Research Center}\\
Patras, Greece\\
ORCID:0000-0002-7517-6997}
\and
\IEEEauthorblockN{Dimitrios Amaxilatis}
\IEEEauthorblockA{\textit{Spark Works Ltd.} \\
Galway, Ireland\\
\\
ORCID:0000-0002-8187-8493}
\and
\IEEEauthorblockN{Michail Giannakos}
\IEEEauthorblockA{\textit{Dpt. of Computer Science} \\
\textit{NTNU}\\
Trondheim, Norway\\
ORCID:0000-0002-8016-6208
}
\and
\IEEEauthorblockN{Georgios Mylonas\thanks{*Corresponding author: Georgios Mylonas, mylonasg@athenarc.gr}\thanks{This work was partially supported by the project MIS 5154714 of the National Recovery and Resilience Plan Greece 2.0 funded by the European Union under the NextGenerationEU Program, and by the Research Council of Norway (RCN) through the AI Centre for the Empowerment of Human Learning (AI LEARN), with a project number: 357493.}}
\IEEEauthorblockA{\textit{Industrial Systems Institute} \\
\textit{Athena Research Center}\\
Patras, Greece\\
ORCID:0000-0003-2128-720X\\
}
}

\IEEEoverridecommandlockouts
\IEEEpubid{\makebox[\columnwidth]{
\hfill} \hspace{\columnsep}\makebox[\columnwidth]{ }}

\maketitle

\copyrightnotice

 
\begin{abstract}
Digital Twins for educational buildings can support sustainability-oriented learning, but their use in schools remains limited. This paper presents a GenAI-based chatbot built on top of an existing Digital Twin for two school buildings in Greece, using real IoT data from environmental sensors and energy meters. The chatbot enables educators to query live and historical building data, compare spaces, and generate ideas for classroom activities through natural language. The system was evaluated in an 80-minute workshop with 17 secondary-school educators, who compared it with an existing web-based dashboard. Results show strong perceived usability and pedagogical value, particularly for inquiry-based learning, hypothesis formation, and interdisciplinary lesson planning. Participants also highlighted limitations related to response speed, data verification, trust, and the continued value of visual dashboards. Overall, the findings suggest that GenAI interfaces can make Digital Twin data more accessible for educational use, provided they are designed with transparency, verification, and pedagogical grounding.

\end{abstract}


\section{Introduction}

To raise awareness of the climate crisis and potential responses, numerous policies and initiatives have emerged worldwide in recent years. Particularly relevant are interventions in the educational sector that combine environmental education with technologies such as IoT and sensor networks. By connecting the physical and digital worlds, these approaches enable the use of real-world data to foster student engagement, creativity, and the integration of sustainability topics into school curricula. Examples such as \cite{gaia-results}, have shown strong potential both for increasing students’ sustainability awareness and for achieving energy savings in participating schools.

At the same time, Digital Twins (DTs) are rapidly becoming popular in an ever-expanding list of application domains, including sustainability in the building sector. In this context, according to \cite{eurostat}, educational buildings in Europe constitute on average 17\% of the non-residential building stock in m$^2$, representing a significant part of the building stock, although numbers can vary significantly between countries. However, even though the importance of such buildings within a city is evident, DT implementations for educational buildings are still few and far between. 

Furthermore, the dizzying progress within LLMs and AI over the past few years has kick-started an ongoing discussion about integrating such technologies in education \cite{ng2025opportunities}. Overall, there are many questions still left unanswered with respect to the use of GenAI-based approaches in education \cite{Giannakos03072025}. Aspects like the financial and energy costs of GenAI, its tendency towards hallucination, and privacy risks further complicate its integration in school environments.

In an attempt to combine the aforementioned aspects, this work discusses the integration of an LLM-based approach utilizing IoT data from the real world, and specifically, environmental and energy data from school buildings, as a means to enable sustainability-oriented activities in an educational setting. The work discussed here is a continuation of our previous efforts reported in \cite{isc2-2025}, where the implementation of a DT for two school buildings was presented. An IoT infrastructure measuring energy consumption and environmental parameters, along with the 3D digitization of the school buildings and the related Web-based user interfaces, resulted in a DT implementation available for use by the educational and research community. 

On top of this implementation, an LLM-based chatbot was built, with the goal of providing a more interactive manner for engaging with the DT and especially with the measurements produced by the sensing infrastructure. Our overall goal was to check whether such a system would open up new possibilities for the educators, in terms of facilitating their access to and use of the data on the one hand, and on the other, enabling them to build educational activities upon the new capabilities on offer. The system was evaluated by \textbf{17} educators of lower secondary school, first via a structured workshop and interviews that took place at the school.  

Based on this motivation, the study is guided by 3 research questions (RQs) addressing the technical feasibility, interface value, and pedagogical potential of the proposed approach:

\begin{itemize}
    \item RQ1: How can an LLM-based conversational interface be implemented on top of a school's DT to provide access to real-world IoT data?
    \item RQ2: What advantages and limitations does a chatbot interface offer compared to conventional Web-based interfaces for exploring school DT data?
    \item RQ3: How do educators perceive the potential of such a system to support the design of educational activities and lesson plans?
\end{itemize}

\section{Previous Related Work}

In this work, we focus on secondary-school contexts, considering both DT and the use of GenAI in educational settings. Existing work on DT in education has largely concentrated on university campuses, including the implementations presented in \cite{cambridge-campus} and \cite{dt-campus-lessons}. These systems typically emphasize real-time monitoring, visualization, and operational decision-making. Related studies have also explored the role of DT in energy efficiency and sustainability assessment, for example, in \cite{dt-sustainability-assessment}.

DT for educational buildings can integrate diverse data sources, including IoT sensors, Building Information Models (BIMs), Building Management Systems (BMSs), and asset or space-management systems \cite{cambridge-campus}. This integration can support improved decision-making and building performance, particularly in areas such as resource management and space utilization, which are central concerns in large campus environments. However, despite their relevance to educational buildings, existing DT implementations have rarely examined how such systems can be used directly for teaching and learning. In particular, the systematic use of school-building DT as educational tools remains underexplored.

Previous work on the utilization of LLMs and GenAI in education, Giannakos et al. \cite{Giannakos03072025} highlight both the promise and the challenges of using GenAI tools in education. While such technologies may support personalization, creativity, and new forms of interaction with learning materials, their adoption also raises important concerns related to efficacy, ethics, privacy, hallucination, overreliance, and pedagogical soundness. These concerns are particularly relevant in school settings, where GenAI systems need to be carefully designed and evaluated before being integrated into classroom practice.

At the same time, recent studies have begun to explore the intersection of GenAI, IoT, and data-rich learning environments. For example, \cite{priya2024harnessing} discusses the potential of combining GenAI with IoT-enhanced education, emphasizing how connected devices and intelligent interfaces can support more interactive and personalized learning experiences. Similarly, \cite{tabuenca2024iot} investigates the use of IoT and generative AI technologies to support urban environmental learning, showing how environmental data can be used as a basis for situated and inquiry-oriented educational activities. In the context of engineering education, \cite{takara2025harnessing} explores the role of GenAI in data science education for civil and environmental engineering, highlighting its potential to help learners interpret complex datasets and engage with domain-specific problems.

More broadly, previous work on multimodal sensor data in education has shown the value of collecting and analyzing data from physical learning environments. For instance, \cite{li2025openmmla} presents OpenMMLA, an IoT-based multimodal data collection toolkit for learning analytics, demonstrating how sensor-based data can support the study and enhancement of learning processes. Such approaches indicate a growing interest in bridging physical environments, data collection infrastructures, and educational analytics. However, most existing work focuses either on sensor and IoT-enhanced learning or GenAI-supported education as separate strands.

Our work builds on these emerging directions by combining real-world IoT data, a school-building DT, and an LLM-based conversational interface in a single educational setting. In contrast to prior work that primarily addresses university-level education, general IoT-enhanced learning, or urban environmental learning, our focus is on secondary-school educators and on the use of school-building data for sustainability-oriented classroom activities. To the best of our knowledge, the systematic exploration of GenAI-supported interaction with a school's DT for this purpose remains underexplored.

\section{Architecture \& Implementation}

Our system implements a conversational natural language interface for querying the IoT sensor deployment, following a serverless architecture hosted entirely on AWS. 
It comprises three logical tiers (Figure~\ref{fig:architecture}): a static web front-end, a serverless back-end with an agentic LLM pipeline, and an IoT platform REST API. The overall system utilizes data provided by a previously developed DT~\cite{isc2-2025} for 2 school buildings, located at the University of Patras, Greece. These buildings house a lower and an upper secondary school, with 46 educators and nearly 390 students.

\begin{figure}
    \centering
    \includegraphics[width=0.48\linewidth]{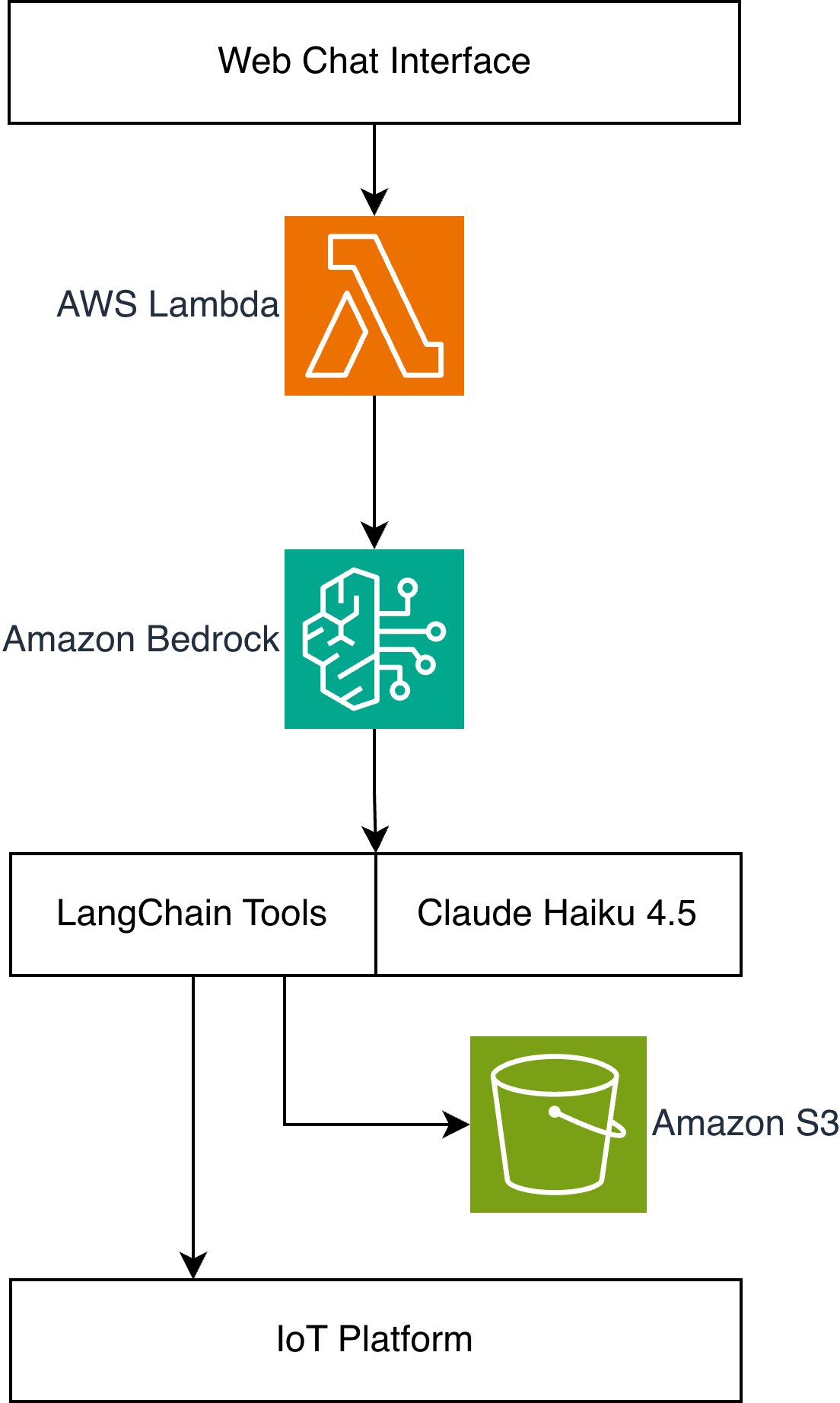}
    \caption{Overview of the system's architecture.}
    \label{fig:architecture}
\end{figure}

\subsection{IoT infrastructure and Digital Twin}

An IoT infrastructure is installed within the 2 school buildings, comprising environmental sensing nodes and power meters. There are 12 environmental nodes (measuring temperature, humidity, noise levels, motion, and luminosity) installed in 12 classrooms (6 in each building), while 4 power meters monitor overall power consumption (2 in each building). The two buildings have also been digitized, and a detailed 3D model is available. Two different Web-based user interfaces (UIs) were provided to the schools: a) a conventional UI, allowing users to retrieve data from the sensors in real time in various formats, as well as retrieve large datasets over long periods of time, and b) a second UI, based on the 3D model, allowing users to see only the latest sensor readings in a ``visual'' manner. In addition to these user interfaces, a set of REST APIs is available for these specific buildings, enabling programmatic access to data collected by the sensing infrastructure. As a result, the implemented infrastructure forms a DT aimed at supporting both educational and research activities. To the best of our knowledge, it represents the first school-building DT implementation of its kind.

\subsection{Front-End (User Interface)}

The chatbot UI (Fig.~\ref{fig:chatbot-ui}) is implemented as a static single-page application. It maintains the dialogue history during each session and forwards user queries, together with the relevant conversation context and access credentials, to the back-end through HTTP requests.

\subsection{Back-End (AWS Lambda)}

The back-end is implemented as a serverless AWS Lambda function. Each request is first authenticated against an Amazon DynamoDB table, ensuring that only authorised users can access the system. Authenticated requests are then processed through an agentic LLM pipeline built with LangChain~\cite{langchain} and deployed via AWS Bedrock. The pipeline uses Anthropic Claude Haiku 4.5 as the main language model, although the model can be configured at deployment time.

The LLM is connected to a set of tools that provide controlled access to the IoT platform. For each user query, the model may invoke one or more tools to retrieve the necessary information before generating a final natural-language response. This loop enables multi-step queries, such as identifying a room, retrieving its sensors, and obtaining current or historical readings, while abstracting away internal identifiers and API details from the end user.

\subsection{Agentic Tool Set}

Access to the IoT platform is mediated through six tools, summarized in Table~\ref{tab:aitools}. These tools allow the LLM to navigate the school--room--sensor hierarchy, retrieve live or historical sensor measurements, and perform name-based searches. By constraining all data access through this tool set, the system supports grounded responses based on the DT data while keeping the interaction conversational for educators.

\begin{table*}[]
    \centering
    \begin{tabular}{|l|l|}
        \hline
        \rowcolor{headblue}
        \textbf{Tool} & \textbf{Purpose} \\
        \hline
        `get\_schools` & Retrieve the list of monitored schools \\
        \hline
        `get\_rooms(school\_uuid)` & Retrieve the rooms within a given school \\
        \hline
        `get\_sensors(room\_uuid)` & Retrieve the sensors installed in a given room \\
        \hline
        `get\_sensor\_reading(sensor\_uuid)` & Retrieve the latest real-time reading for a sensor \\
        \hline
        `get\_sensor\_reading\_history(sensor\_uuid, from, to)` & Retrieve hourly aggregated historical readings over a time range \\
        \hline
        `search\_by\_name(query, type)` & Full-text search across school and room names \\
        \hline
    \end{tabular}
    \caption{List of tools giving data access to the IoT platform.}
    \label{tab:aitools}
\end{table*}

\subsubsection{IoT Platform API}

The IoT REST API is protected with OAuth 2.0 (Resource Owner Password Credentials grant). The Lambda function obtains and caches a bearer token for the lifetime of the execution environment. Structural metadata (school lists, room lists, sensor lists) is additionally cached in an \textbf{Amazon S3} bucket as a JSON document, avoiding redundant API calls for data that changes infrequently.

\subsubsection{Infrastructure}

The entire back-end is defined as infrastructure-as-code using an \textbf{AWS SAM} CloudFormation template, which provisions the Lambda function, the DynamoDB passwords table, the S3 static-data cache bucket, and the associated IAM roles and policies. The stack is registered with AWS Service Catalog AppRegistry for cost attribution. The LLM provider and model are configurable at deploy time via CloudFormation parameters.

\section{Envisioned applications}

Previous activities targeting sustainability within the classroom based on IoT data have produced promising results \cite{gaia-results}, specifically in terms of promoting sustainability awareness and energy saving. Moreover, there are multiple educational contexts where DTs like the one utilized in this study could fit in the curriculum of primary and secondary schools. 

At its most immediate level, the system allows students and teachers to query their physical environment in plain language. A student can ask ``What is the temperature in the Computer Lab?'' and receive a live reading from the room's sensors, or ask ``Show me all the sensors in the Physics classroom.'' and get an overview of what is being monitored and where. A teacher preparing a lesson can ask, "Which classrooms are in the High School?" to orient themselves within the building's DT. It can also provide users with historical data by asking ``What was the mean temperature in the computer Lab this week?''. Such interactions, while seemingly simple, mark an important shift: building data moves from technical datasets or dashboards, typically accessible only to facility managers or expert users, to a conversational and explorable resource that can become meaningful to the wider school community.

\begin{figure}
    \centering
    \includegraphics[width=0.95\linewidth]{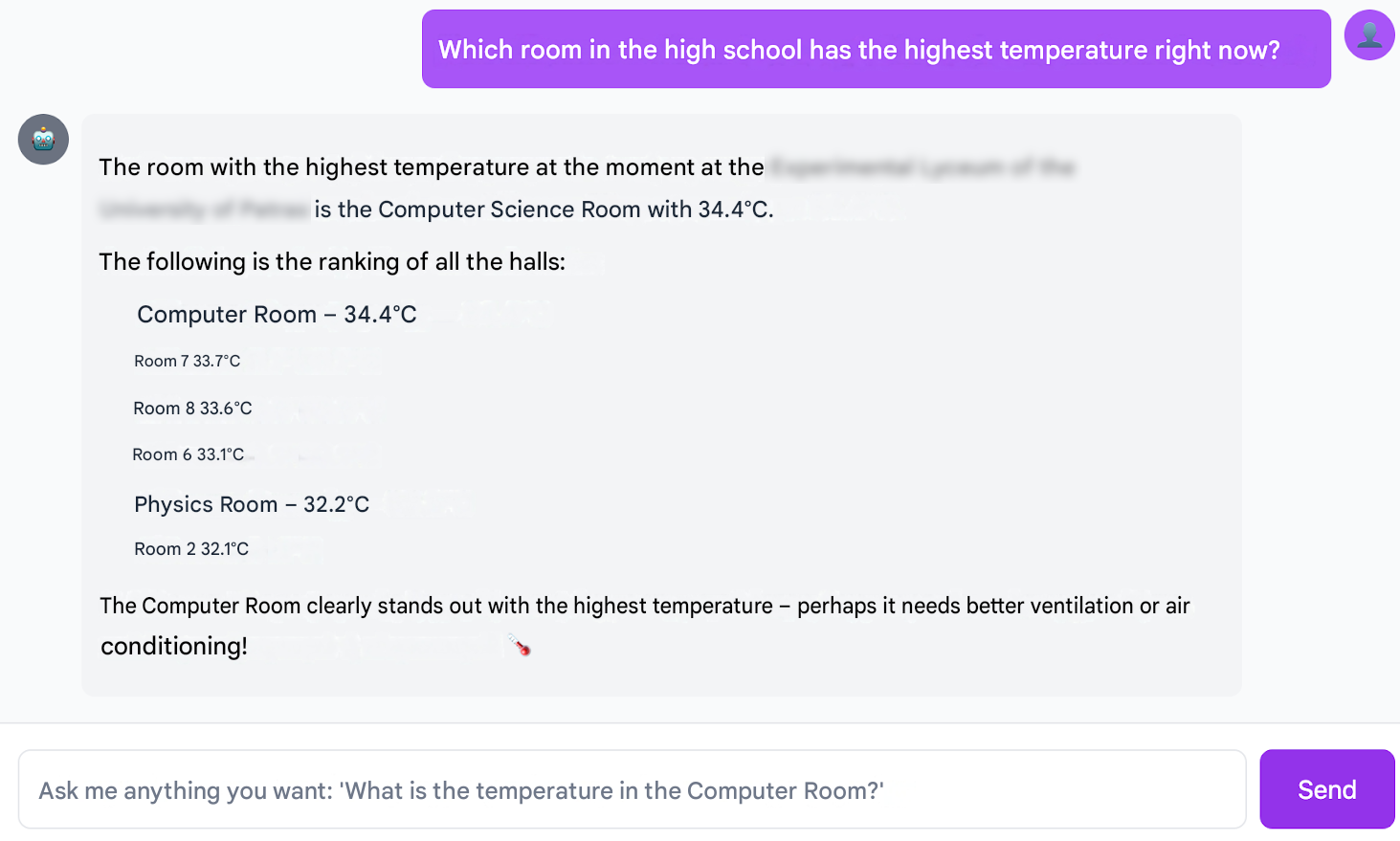}
    \caption{Example of the user interface, with the user asking for the room with the highest temperature in the school.}
    \label{fig:chatbot-ui}
\end{figure}

In more complex scenarios, the system can help users identify patterns and make correlations between the data, while also making suggestions to the users regarding potential educational uses. For example, the user could attempt a comparison between the temperature in similar rooms in the two schools, and the system could respond that there appear to be significant differences, prompting the user to trigger a process to investigate what the reason for this situation might be. Similarly, if a classroom's temperature keeps rising throughout the week, rather than simply logging the anomaly, the AI assistant surfaces this to the user: ``Temperature levels in your classroom have risen since last week. What do you think is causing this, and what might happen if this continues?''. This transforms an ordinary sensor reading into a vivid conversation, prompting users to think and connect human activity with ventilation and other concepts spanning from biology to physics.

\section{Evaluation - Results}

\subsection{Chatbot Performance Evaluation}
During the development of our AI assistant, we evaluated several LLM models before settling on Claude. The models tested included Qwen (30B), Llama 3.2 (1B/3B), Gemma 4 (2b/4b), Claude Haiku 4.5, and Claude Opus 4.6. Our use case required multi-step reasoning across sequential API calls, which could not be resolved in a single interaction. Smaller models, such as Llama Gemma and Qwen, consistently failed, losing context across calls and producing incomplete replies due to their constraints. Larger models showed a clear qualitative improvement, though we observed that even such models would at times hallucinate a plausible answer, rather than issue API calls required to retrieve data from the system.

As a remedy to this situation, LangChain tools integration can allow models to identify whether they lack sufficient information and issue additional API calls until a grounded response can be formed, minimizing hallucination risk. Smaller models could not leverage this, as their capacity was insufficient to manage tool orchestration alongside reasoning. Claude, in contrast, handled chained tool use reliably, producing accurate and coherent responses across multi-step queries. It was therefore selected as the default model, as it offered the best balance between reasoning capability, response quality, and reliability for the intended end-user scenarios.

\subsection{User Evaluation Setup}

An evaluation workshop was conducted with 17 educators of various disciplines from the lower secondary school participating in the study. The workshop lasted 80 minutes and was designed to facilitate a direct comparison between the "legacy" web-based dashboard and the chatbot. All participants were provided with an informed consent document, compliant with GDPR. The session was structured into 3 phases:

\begin{itemize}
    \item Introduction \& Interface Walkthrough (20 minutes): An overview of the school's sensing infrastructure, followed by a demonstration of the available user interfaces.
    \item Comparative Hands-on Tasks (40 minutes): Participants engaged in 2 activities. The first required data retrieval and correlation using the web UI, while the second focused on insight and correlation using the chatbot.
    \item Synthesis and Data Collection (20 minutes): The session concluded with a collective reflection and a discussion regarding the integration of the chatbot in educational activities and its potential role in crafting new lesson plans, followed by the administration of the study instruments.
\end{itemize}

The evaluation comprised the filling of a questionnaire by participants. Adapted from a coding-pedagogy framework, the questionnaire gathered quantitative and qualitative data across several dimensions:

\begin{itemize}
    \item Usability and Intuitiveness: evaluating the functionality of the conversational interface.
    \item Effectiveness: assessing how well the tool helped educators solve specific data challenges.
    \item Critical Thinking: determining whether the tool encouraged deeper reflection on the school's environment.
    \item Comparative Value: direct comparison of the chatbot against other methods.
\end{itemize}

\subsection{Results}

The first question (Q1) captured participants' self-reported familiarity with ICT using a 5-point Likert scale. The remaining 18 questionnaire items were also measured on 5-point Likert scales and were grouped into four thematic dimensions: Usability (Q2--Q6), Trust \& Transparency (Q7--Q11), Pedagogical Value (Q12--Q16), and Limitations \& Readiness (Q17--Q19). Responses were collected from all 17 participants at the end of the evaluation workshop.

The analysis combined descriptive and non-parametric statistics. Wilcoxon signed-rank tests were used to compare responses against the neutral midpoint of 3, Mann-Whitney U tests were used for group comparisons, and Spearman's $\rho$ was used to examine correlations between items. Given the sample size of $n = 17$, correlations were interpreted cautiously; coefficients of approximately $|r| > 0.48$ correspond to $p < 0.05$.

\subsubsection{Overall Ratings}

Table~\ref{tab:means} reports the mean and standard deviation for each questionnaire item. Fourteen of the eighteen items received mean scores above 4, indicating strong agreement across the majority of the sample. Q2 (ease of use) recorded the highest score ($M = 4.69$, $SD = 0.79$), while Q9 (trust in chatbot answers, $M = 3.76$) and Q7 (pattern detection, $M = 3.82$) were the most modestly rated positive items, both still sitting comfortably above the neutral midpoint. Items Q17--Q19, which capture perceived limitations and integration readiness (new challenges, preference for the web UI, and security concerns), were closer to the neutral midpoint ($M \approx 3.5$--$3.8$), reflecting a degree of uncertainty rather than outright resistance. Overall, the results indicate that usability and pedagogical-value items were rated more positively than limitation-related items, with no item falling below the neutral midpoint.

\begin{table}[H]
  \centering
  \caption{Per-item means and standard deviations ($n = 17$).}
  \label{tab:means}
  \small
  \setlength{\tabcolsep}{4pt}
  \begin{tabular}{llcc}
    \toprule
    \rowcolor{headblue}
    \textbf{Item} & \textbf{Description} & \textbf{M} & \textbf{SD} \\
    \midrule
    Q1  & Familiarity with ICT    & \cellcolor{cellgreen} 4.18 & 1.01 \\
    Q2  & Ease of use             & \cellcolor{cellgreen} 4.69 & 0.79 \\
    Q3  & More intuitive          & \cellcolor{cellgreen} 4.31 & 0.70 \\
    Q4  & Fewer steps             & \cellcolor{cellgreen} 4.29 & 0.69 \\
    Q5  & Comparing data          & \cellcolor{cellgreen} 4.24 & 0.83 \\
    Q6  & Clear answers           & \cellcolor{cellgreen} 4.18 & 0.73 \\
    Q7  & Pattern detection       & \cellcolor{cellgray} 3.82  & 0.81 \\
    Q8  & Deeper thinking         & \cellcolor{cellgreen} 4.29 & 0.77 \\
    Q9  & Trust                   & \cellcolor{cellgray} 3.76  & 1.03 \\
    Q10 & Transparency            & \cellcolor{cellgreen} 4.06 & 1.03 \\
    Q11 & Verification            & \cellcolor{cellgreen} 4.12 & 0.78 \\
    Q12 & Student research        & \cellcolor{cellgreen} 4.41 & 0.80 \\
    Q13 & Hypotheses              & \cellcolor{cellgreen} 4.29 & 0.77 \\
    Q14 & Accessibility           & \cellcolor{cellgreen} 4.29 & 0.99 \\
    Q15 & Integration in teaching & \cellcolor{cellgreen} 4.24 & 0.97 \\
    Q16 & Future use              & \cellcolor{cellgreen} 4.41 & 1.00 \\
    Q17 & New challenges          & \cellcolor{cellgray} 3.76  & 0.83 \\
    Q18 & Web $>$ chatbot         & \cellcolor{cellgray} 3.53  & 0.94 \\
    Q19 & Security concerns       & \cellcolor{cellgray} 3.62  & 1.15 \\
    \bottomrule
    \multicolumn{4}{l}{\footnotesize \textcolor{cellgreen}{\rule{2mm}{2mm}}$M \geq 4$ \quad
                        \textcolor{cellgray}{\rule{2mm}{2mm}}$3 \leq M < 4$}
  \end{tabular}
\end{table}

\subsubsection{Section-Level Composite Scores}

Aggregating items into the four thematic sections (Table~\ref{tab:composite}, Figure~\ref{fig:composite}) reinforces the item-level picture. Pedagogical Value achieved the highest composite mean ($M = 4.33$, $p = 0.001$ vs. neutral), followed
closely by Usability ($M = 4.32$, $p < 0.001$) and Trust \& Transparency ($M = 4.01$, $p = 0.001$). Limitations \& Readiness scored near the neutral midpoint ($M = 3.63$, $p = 0.008$), suggesting mild rather than strong concerns.

Splitting participants by self-reported ICT familiarity (high: $Q1 \geq 4$, $n = 14$; low: $Q1 \leq 3$, $n = 3$) reveals a statistically significant difference only in Usability ($p = 0.026$) and Pedagogical Value ($p = 0.032$), with high-familiarity teachers rating Usability higher ($M = 4.43$ vs. $3.83$) while low-familiarity teachers actually rated Pedagogical Value higher ($M = 4.87$ vs. $4.21$). The low group is small and these differences should be interpreted cautiously, but the pattern suggests that even educators with limited technology experience perceived clear pedagogical merit in the tool.

\begin{table}[H]
  \centering
  \caption{Composite scores by section and ICT familiarity group.}
  \label{tab:composite}
  \small
  \setlength{\tabcolsep}{3.5pt}
  \resizebox{\columnwidth}{!}{
  \begin{tabular}{lcccccc}
    \toprule
    \rowcolor{headblue}
    \textbf{Section} & \textbf{M} & \textbf{SD}
      & \textbf{M\textsubscript{High}} & \textbf{M\textsubscript{Low}}
      & \textbf{p\textsubscript{MW}} & \textbf{p\textsubscript{Wilc.}} \\
    \midrule
    C.1 Usability          & 4.32 & 0.43 & 4.43 & 3.83 & $0.026^{*}$ & $<0.001^{*}$ \\
    C.2 Trust \& Transp.   & 4.01 & 0.59 & 4.07 & 3.73 & 0.337       & $0.001^{*}$  \\
    C.3 Pedagogical Value  & 4.33 & 0.77 & 4.21 & 4.87 & $0.032^{*}$ & $0.001^{*}$  \\
    C.4 Limitations        & 3.63 & 0.80 & 3.55 & 4.00 & 0.522       & $0.008^{*}$  \\
    \bottomrule
    \multicolumn{7}{l}{\footnotesize $^{*}p < 0.05$; MW = Mann-Whitney U; Wilc. = Wilcoxon vs. neutral 3.}
  \end{tabular}
  }
\end{table}

\begin{figure}[H]
  \centering
  \includegraphics[width=\columnwidth]{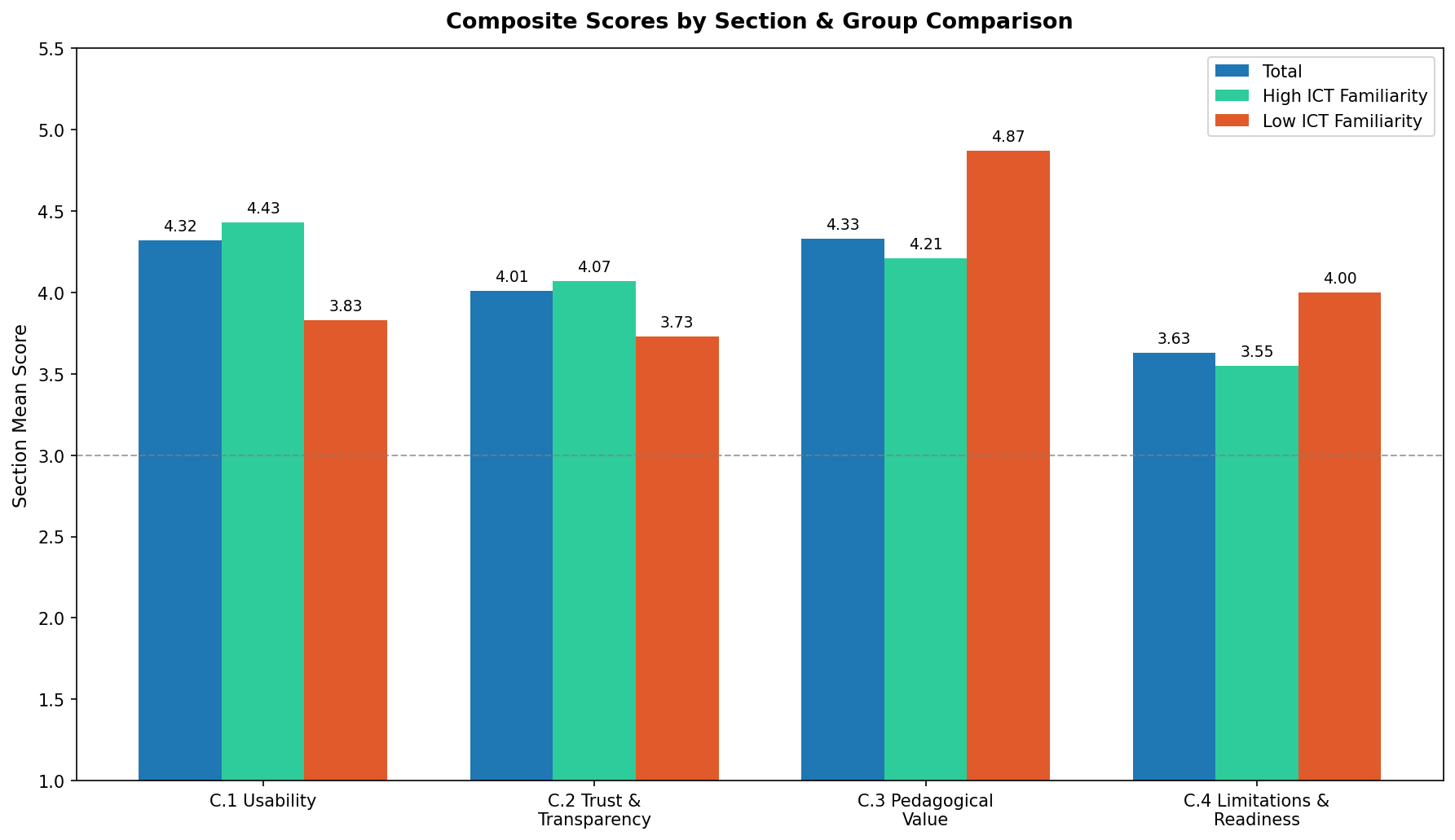}
  \caption{Composite section scores by ICT familiarity group.
           Dashed line marks the neutral midpoint.}
  \label{fig:composite}
\end{figure}

\subsubsection{Effect of ICT Familiarity}

To assess whether prior technology experience moderated
attitudes toward the chatbot, we computed Spearman correlations
between Q1 (self-reported ICT familiarity) and each item
(Figure~\ref{fig:spearman_q1}).
No item reached statistical significance at $p < 0.05$
($|r| > 0.48$ required for $n = 17$), confirming that the
chatbot was evaluated positively irrespective of participants'
technology background. The only noteworthy signal was a
marginal negative trend at Q15 (integration in teaching,
$r = -0.475$, $p = 0.054$): more experienced educators were
slightly less optimistic about integrating the tool into
lessons, a finding that warrants exploration with a larger sample.

\begin{figure}[H]
  \centering
  \includegraphics[width=\columnwidth]{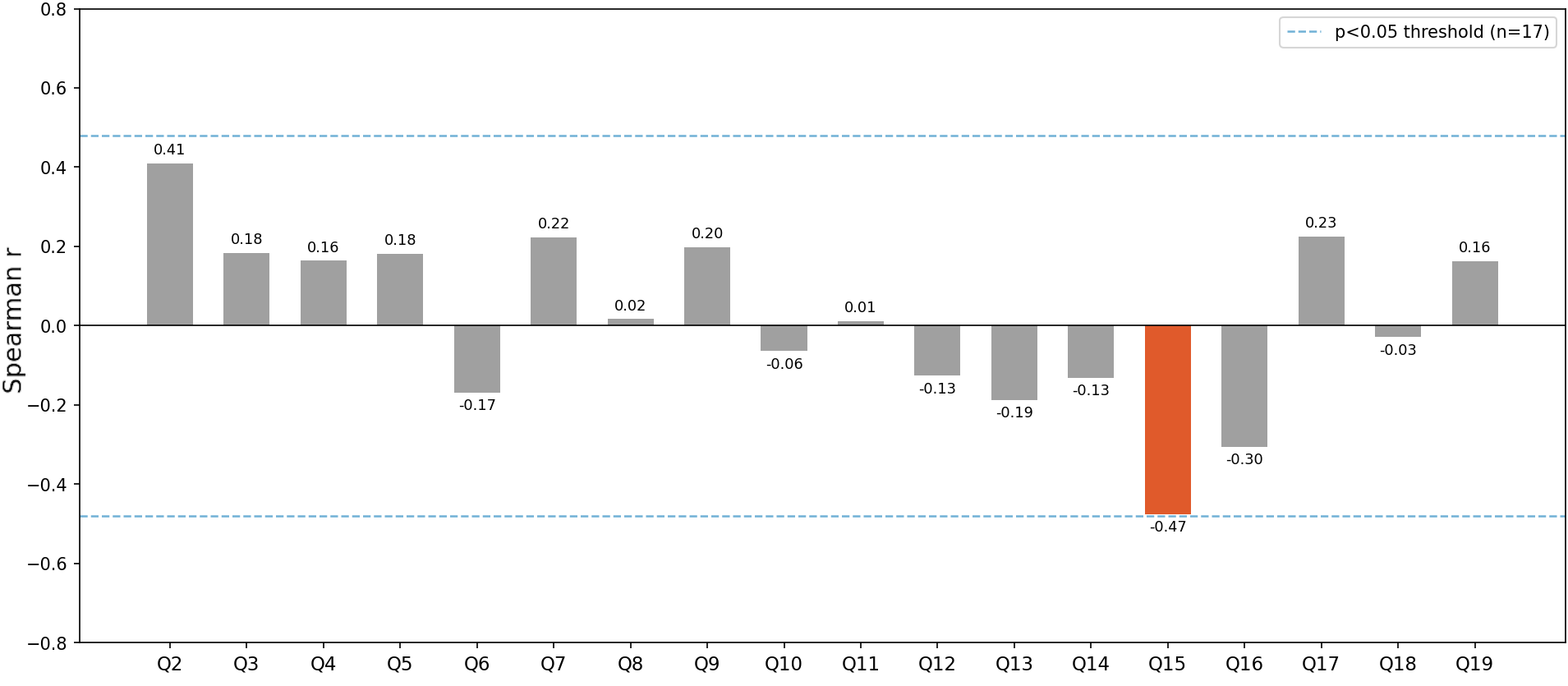}
  \caption{Spearman $r$ between ICT familiarity (Q1) and each item.
           Dashed lines mark the $p{<}0.05$ significance threshold
           ($|r| = 0.48$, $n = 17$). No item reaches significance.}
  \label{fig:spearman_q1}
\end{figure}

\subsubsection{Inter-Item Correlations}

Pairwise Spearman correlations revealed nine pairs with
$|r| \geq 0.6$ and $p < 0.05$ (Table~\ref{tab:sig_pairs},
Figure~\ref{fig:heatmap}). The strongest association was
between Q12 (student research) and Q13 (hypothesis formation,
$r = 0.825$), suggesting that teachers perceive these as a
unified pedagogical affordance. Similarly, Q8 (deeper thinking)
and Q14 (accessibility) co-varied strongly ($r = 0.776$),
pointing to a latent construct linking cognitive depth with
ease of access to information. Within the usability cluster,
Q2 (ease of use), Q4 (fewer steps), and Q7 (pattern detection)
formed a tight triad ($r = 0.62$--$0.69$), indicating that
perceived ease of interaction generalises to perceived
analytical utility. Notably, Q18 (preference for web UI over
chatbot) correlated negatively with both Q2 ($r = -0.606$) and
Q7 ($r = -0.608$): participants who found the chatbot easy
and analytically useful were least inclined to revert to the
legacy interface.

\begin{table}[H]
  \centering
  \caption{Significant inter-item Spearman pairs ($|r|\geq 0.6$, $p<0.05$).}
  \label{tab:sig_pairs}
  \small
  \begin{tabularx}{\columnwidth}{lccc X}
    \toprule
    \rowcolor{headblue}
    \textbf{Pair} & \textbf{r} & \textbf{p} & \textbf{Dir.} & \textbf{Note} \\
    \midrule
    Q12$\leftrightarrow$Q13 & 0.825 & $<$0.001 & + & Research $\leftrightarrow$ Hypotheses \\
    Q8$\leftrightarrow$Q14  & 0.776 & $<$0.001 & + & Thinking $\leftrightarrow$ Accessibility \\
    Q2$\leftrightarrow$Q7   & 0.685 & 0.003    & + & Ease $\leftrightarrow$ Pattern detection \\
    Q18$\leftrightarrow$Q19 & 0.664 & 0.005    & + & Limitations converge \\
    Q10$\leftrightarrow$Q11 & 0.662 & 0.004    & + & Transparency $\leftrightarrow$ Verification \\
    Q2$\leftrightarrow$Q4   & 0.621 & 0.010    & + & Ease $\leftrightarrow$ Fewer steps \\
    Q4$\leftrightarrow$Q7   & 0.619 & 0.008    & + & Steps $\leftrightarrow$ Pattern detection \\
    Q2$\leftrightarrow$Q18  & $-$0.606 & 0.013 & $-$ & Ease $\leftrightarrow$ Web preference \\
    Q7$\leftrightarrow$Q18  & $-$0.608 & 0.010 & $-$ & Detection $\leftrightarrow$ Web preference \\
    \bottomrule
  \end{tabularx}
\end{table}

\begin{figure}[H]
  \centering
  \includegraphics[width=\columnwidth]{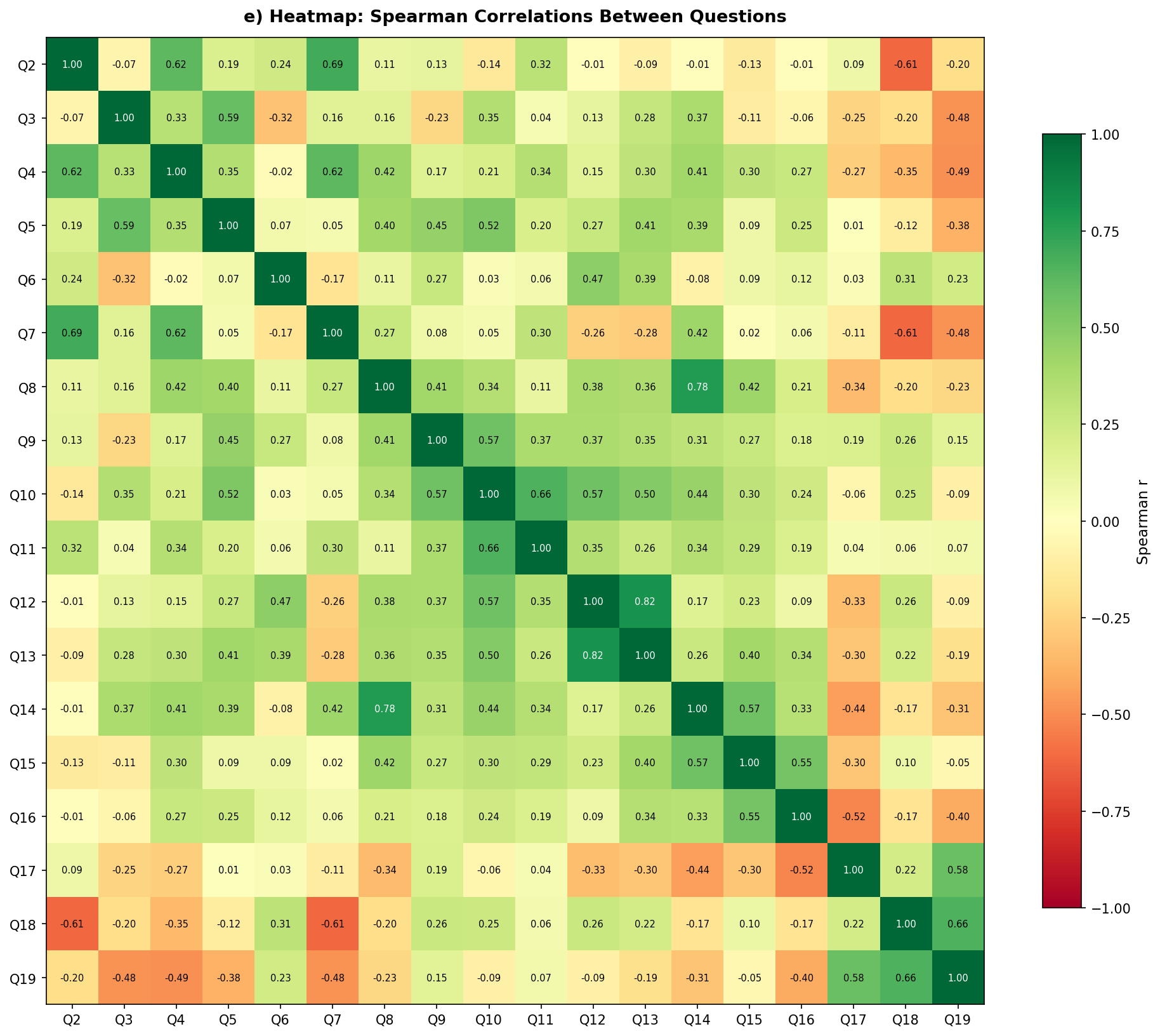}
  \caption{Spearman correlation heatmap across all items.}
  \label{fig:heatmap}
\end{figure}

\section{Discussion}

With respect to RQ1, the results suggest that tool-mediated LLM orchestration is important for producing reliable responses over DT data. In our implementation, LangChain tools enabled the model to access the IoT platform through controlled API calls rather than relying on unsupported generation. This was particularly important for multi-step queries, where the system first had to resolve the relevant school, room, and sensor before retrieving live or historical measurements. Smaller locally hosted models were not consistently reliable for this type of chained interaction, leading us to adopt Claude as the default cloud-based model. Nevertheless, this choice introduces practical constraints related to data protection, school network policies, cost, and dependence on external services. Future work should therefore examine whether larger locally deployed models, supported by appropriate hardware, can offer comparable reliability under stricter institutional requirements.

Regarding RQ2, the questionnaire and free-text responses indicate that participants generally perceived the chatbot as easier and more intuitive than the conventional web-based interface. They valued its ability to support natural-language exploration, comparisons between measurements, correlations across spaces or data types, and more reflective engagement with the sensing data. These findings suggest that the chatbot is particularly useful for exploratory and inquiry-oriented scenarios, where educators or students may not know in advance which dashboard view or dataset they need. At the same time, the legacy interface retained clear advantages for specific tasks. Participants noted that it could provide faster access to individual sensor readings, immediate chart generation, and better visual monitoring of data values. It was also seen as useful for independently verifying the chatbot's answers. Thus, rather than replacing the dashboard, the chatbot should be viewed as a complementary interface that broadens access to DT data while still benefiting from visual and verification-oriented tools.

With respect to RQ3, participants perceived the system as promising for supporting sustainability-oriented educational activities and lesson-plan design. Questionnaire responses were positive regarding student inquiry, hypothesis formation, integration in teaching, and future use. Free-text comments further suggested potential applications across diverse subjects, including foreign languages, CLIL-based activities, theoretical courses, and interdisciplinary projects. Several educators also appreciated the chatbot's ability to propose subject-specific activities, evaluation ideas, and personalized lesson-plan suggestions based on real school data. However, participants also emphasized that such uses should be framed carefully. They highlighted the need for safeguards, attention to personal data, improved response speed, and confidence in data quality. Importantly, they stressed that students should not trust chatbot outputs blindly, but should use them as a starting point for critical thinking, verification, and discussion.

\subsection{Limitations}

The study included here does not reflect on long-term use of the chatbot, i.e., beyond the workshop conducted, thus limiting the generalizability of the results in terms of user retention and long-term effects. In addition, the sample of the educators was not balanced in terms of gender. We have also not taken into consideration the energy resources associated with the chatbot's operation, nor could we communicate them to the educators, since AWS did not provide this data; in the context of educational activities for sustainability, such data could be important to the system's end users, in order to have a more complete picture concerning its use and related resources.

\section{Conclusion}

DTs can contribute to more sustainable and resilient cities, while school-building DTs create additional opportunities for teaching and learning around real-world environmental data. At the same time, GenAI introduces new possibilities for making such data more accessible, but also raises important questions regarding reliability, transparency, and appropriate use in educational settings. In this work, we investigated a GenAI-based conversational interface built on top of a DT/IoT infrastructure deployed in two school buildings, with the aim of supporting access to sensing data and enabling sustainability-oriented educational activities.

The system was evaluated through a workshop with 17 educators, who compared the chatbot with a conventional web-based interface and reflected on its potential educational uses. With respect to RQ1, our findings suggest that reliable interaction with DT data requires tool-mediated LLM orchestration and sufficient model reasoning capacity. In our case, smaller locally deployed models were not reliable enough for multi-step data retrieval, while Claude provided more consistent performance through chained tool use. With respect to RQ2, participants generally perceived the chatbot as easier and more intuitive than the conventional interface, particularly for asking questions, comparing measurements, and exploring correlations. However, the web-based interface remained useful for tasks requiring quick visual inspection, charting, and independent verification of sensor values. Regarding RQ3, educators perceived the system as promising for supporting learning activities, including inquiry-based learning, hypothesis formation, interdisciplinary projects, and lesson-plan design. At the same time, their responses highlighted the need for careful integration in classroom practice, especially with respect to trust, data verification, response speed, privacy, and the importance of encouraging students to critically assess chatbot outputs rather than accept them uncritically.

Future work will focus on a more extensive evaluation with educators from additional schools and educational levels, as well as with student groups. We also plan to investigate tighter integration with other components of the DT, including the 3D building models, and to explore applications beyond classroom activities, such as building operation, maintenance, and energy-awareness interventions.

\section*{Acknowledgments}

The authors would like to thank the staff of the Experimental Lower and Upper Secondary Schools of the University of Patras for their collaboration towards this study.

\bibliographystyle{IEEEtran}
\bibliography{bibliography}

@article{ng2025opportunities,
  title={Opportunities, challenges and school strategies for integrating generative AI in education},
  author={Ng, Davy Tsz Kit and Chan, Eagle Kai Chi and Lo, Chung Kwan},
  journal={Computers and Education: Artificial Intelligence},
  volume={8},
  pages={100373},
  year={2025},
  publisher={Elsevier}
}

@inproceedings{takara2025harnessing,
  title={Harnessing the power of GenAI: A new era for data science education for civil and environmental engineering},
  author={Takara, Matthew Yukio and Ozis, Fethiye and Pensky, Allison E Connell},
  booktitle={2025 ASEE Annual Conference \& Exposition},
  year={2025}
}

@inproceedings{li2025openmmla,
  title={OpenMMLA: An IoT-based multimodal data collection toolkit for learning analytics},
  author={Li, Zaibei and Yamaguchi, Shunpei and Spikol, Daniel},
  booktitle={Proceedings of the 15th International Learning Analytics and Knowledge Conference},
  pages={872--879},
  year={2025}
}

@inproceedings{priya2024harnessing,
  title={Harnessing Generative AI for IoT Enhanced Education},
  author={Priya, KM and Soundarya, P and Kungumaraj, E and Bhagyarathi, P and Subathra, P and Varghese, Susmi Mariam},
  booktitle={2024 IEEE International Conference on Signal Processing, Informatics, Communication and Energy Systems (SPICES)},
  pages={1--6},
  year={2024},
  organization={IEEE}
}

@inproceedings{tabuenca2024iot,
  title={IoT and generative AI technologies to support urban environmental learning},
  author={Tabuenca, Bernardo and Mart{\'\i}n, Sergio and Greller, Wolfgang and Tillmann, Alexander and Uche-Soria, Manuel and Castro, Manuel and Tovar, Edmundo and Rodr{\'\i}guez-Artacho, Miguel},
  booktitle={2024 IEEE global engineering education conference (EDUCON)},
  pages={1--4},
  year={2024},
  organization={IEEE}
}

@ARTICLE{gaia-results,
  author={Mylonas, Georgios and Amaxilatis, Dimitrios and Paganelli, Federica and Chatzigiannakis, Ioannis and Koulouris, Pavlos and Falkouskaya, Yelizaveta and Markopoulos, Panos},
  journal={IEEE Internet of Things Journal}, 
  title={{Results From a Large-Scale IoT-Based Intervention for Energy Saving and Sustainability Awareness via Behavior Change in 25 K-12 Schools in Europe}}, 
  year={2025},
  volume={12},
  number={12},
  pages={19308-19325},
  doi={10.1109/JIOT.2025.3544652}}

@book{eurostat,
author = {Economidou, Marina and Atanasiu, Bogdan and Staniaszek, Dan and Maio, Joana and Nolte, Ingeborg and Rapf, Oliver and Laustsen, Jens and Ruyssevelt, Paul and Strong, David and Zinetti, Silvia},
year = {2011},
month = {10},
title = {Europe's buildings under the microscope. A country-by-country review of the energy performance of buildings},
isbn = {9789491143014},
}

@article{cambridge-campus,
author = {Lu, Qiuchen and Parlikad, Ajith Kumar and Woodall, Philip and Xie, Xiang and Liang, Zhenglin and Konstantinou, Eirini and Heaton, James and Schooling, Jennifer},
year = {2019},
month = {10},
pages = {},
title = {Developing a dynamic digital twin at building and city levels: A case study of the West Cambridge campus},
volume = {36},
journal = {Journal of Management in Engineering},
doi = {10.1061/(ASCE)ME.1943-5479.0000763}
}

@ARTICLE{dt-campus-lessons,
  author={Roda-Sanchez, Luis and Cirillo, Flavio and Solmaz, Gürkan and Jacobs, Tobias and Garrido-Hidalgo, Celia and Olivares, Teresa and Kovacs, Ernö},
  journal={IEEE Internet of Things Journal}, 
  title={Building a Smart Campus Digital Twin: System, Analytics, and Lessons Learned From a Real-World Project}, 
  year={2024},
  volume={11},
  number={3},
  pages={4614-4627},
  doi={10.1109/JIOT.2023.3300447}}

@Article{dt-sustainability-assessment,
AUTHOR = {Tagliabue, Lavinia Chiara and Cecconi, Fulvio Re and Maltese, Sebastiano and Rinaldi, Stefano and Ciribini, Angelo Luigi Camillo and Flammini, Alessandra},
TITLE = {Leveraging Digital Twin for Sustainability Assessment of an Educational Building},
JOURNAL = {Sustainability},
VOLUME = {13},
YEAR = {2021},
NUMBER = {2},
ARTICLE-NUMBER = {480},
URL = {https://www.mdpi.com/2071-1050/13/2/480},
ISSN = {2071-1050},
DOI = {10.3390/su13020480}
}

@article{Giannakos03072025,
author = {Michail Giannakos and Roger Azevedo and Peter Brusilovsky and Mutlu Cukurova and Yannis Dimitriadis and Davinia Hernandez-Leo and Sanna Järvelä and Manolis Mavrikis and Bart Rienties},
title = {The promise and challenges of generative AI in education},
journal = {Behaviour \& Information Technology},
volume = {44},
number = {11},
pages = {2518--2544},
year = {2025},
publisher = {Taylor \& Francis},
doi = {10.1080/0144929X.2024.2394886},
URL = {https://doi.org/10.1080/0144929X.2024.2394886},
eprint = {https://doi.org/10.1080/0144929X.2024.2394886}
}

@INPROCEEDINGS{isc2-2025,
  author={Mylonas, Georgios and Koutsoudis, Anestis and Amaxilatis, Dimitrios and Arnaoutoglou, Fotis and Kalogeras, Georgios and Pistofidis, Petros and Fraile, Lidia Pocero and Koulamas, Christos and Kalogeras, Athanasios},
  booktitle={2025 IEEE International Smart Cities Conference (ISC2)}, 
  title={Implementing a Digital Twin for a Secondary School Building Complex in Greece}, 
  year={2025},
  volume={},
  number={},
  pages={1-6},
  doi={10.1109/ISC266238.2025.11293266}}

@misc{langchain,
 key = "langchain",
 title = {{LangChain, the agent engineering framework, Github Repository}, https://github.com/langchain-ai/langchain},
 year = "accessed May 2026",
}

\end{document}